\documentstyle[aps,epsf]{revtex}

\begin{document}

\draft

\title{Modern meson--exchange potential and
       superfluid neutron star crust matter}

\author{\O.\ Elgar\o y, L.\ Engvik and E.\ Osnes}

\address{Department of Physics, University of Oslo, N--0316 Oslo, Norway}

\author{F.\ V.\ De Blasio, M.\ Hjorth--Jensen and G.\ Lazzari}

\address{ECT*, European Centre for Theoretical Studies in
         Nuclear Physics and Related Areas, \\
         Strada delle Tabarelle 286, I-38050 Villazzano (Trento), Italy}

\maketitle

\begin{abstract}
    In this work we study properties of neutron star crusts, where
    matter is expected to consist of
    nuclei surrounded by superfluid
    neutrons and a homogeneous background of relativistic electrons.
    The nuclei are disposed in a Coulomb lattice, and it is
    believed that the structure of the lattice
    influences considerably
    the specific heat
    of the neutronic matter inside the crust of a neutron star.
    Using a modern meson--exchange potential in the framework of
    a local--density approximation we calculate the neutronic
    specific heat accounting for various shapes of
    the Coulomb lattice, from spherical
    to non--spherical nuclear shapes. We find  that a realistic
    nucleon--nucleon  potential
    leads to a significant increase in the neutronic specific heat
    with respect to
    that
    obtained assuming a uniform neutron distribution.
    The increase is largest for the
    non--spherical phase of the crust. These results may have
    consequences for the thermal history of young neutron stars.
\end{abstract}

\pacs{PACS number(s): 97.60.Jd 21.65.+f 74.25.Bt }

\twocolumn

The observation of thermal emission from the surface of a neutron star is
a powerful tool by which one can obtain information about the state of
matter inside the star. It has been shown that the time needed for a
temperature drop in the core to affect the surface temperature should
depend on the thickness of the crust and on its thermal properties, such
as the total specific heat \cite{bkpp88}, which is strongly influenced
by the superfluid state of matter inside the crust.

It has recently been proposed that the
Coulomb--lattice structure of a neutron
star  crust may influence significantly  the thermodynamical
properties  of the superfluid neutron gas \cite{bdllp94}.
The authors of Refs.
\cite{oyamatsu93,lrp93,prl95,pr95}  have  proposed that in the crust
of a neutron star non--spherical nuclear shapes could be present at
densities ranging from $\rho=1.0\times 10^{14}$ gcm$^{-3}$ to
$\rho=1.5\times 10^{14}$ gcm$^{-3}$, a density
region which represents about $20\%$ of the whole crust. The
saturation density of nuclear matter is $\rho_0 =2.8\times 10^{14}$
gcm$^{-3}$.
These unusual
shapes  are supposed \cite{lrp93,prl95}
to be disposed in a Coulomb lattice embedded in an
almost uniform background of relativistic electrons. According to the
fact that the neutron drip point is supposed to occur at lower density
($\rho\sim 4.3 \times 10^{11}$ gcm$^{-3}$),
 and considering the characteristics of the nuclear force in
this density range, we expect  these
unusual nuclear shapes to be
surrounded by a gas of superfluid neutrons.

In the present paper we follow essentially Ref.\
\cite{bdllp94}, however, we differ in
using one of the
modern meson--exchange potentials of the Bonn group \cite{mac89}
in evaluating the neutron pairing energy gap. This potential
is expected to give, contrary to the effective one used in Ref.\
\cite{bdllp94}, a more realistic estimate of
the gaps in the region inside
the nuclear cluster.

In the following  we treat the crust--lattice in the
Wigner--Seitz
approximation,
dividing the Coulomb lattice into unit cells
of appropriate shape (cylindrical, planar and spherical), containing
a nucleus surrounded by a gas of superfluid neutrons.
Details of this system have been worked out within the framework
of Thomas--Fermi calculations with different energy density
functionals \cite{oyamatsu93}. The neutron and proton density profiles
are obtained from a recent parametrization of Oyamatsu \cite{oyamatsu93}
\begin{equation}
      \rho_i(r)=(\rho_i^{in}-\rho^{out}_i)\left[1-\Bigl({r\over
       R_i}\Bigr)^{t_i}\right]^{k_i}+\rho_i^{out},
\end{equation}
for $r<R_i$ and
\begin{equation}
      \rho_i(r)=\rho_i^{out},
\end{equation}
for $r>R_i$, where $i=p,n$ represent protons and neutrons respectively,
with
$\rho^{out}_p$ taken to be zero.
Following Oyamatsu \cite{oyamatsu93} the spherical shape is expected
for densities $\rho<0.35\rho_0$ (about $80\%$ of the whole crust).
The cylindrical region is supposed to appear for densities $\rho$ between
$0.35\rho_0$ and $0.46\rho_0$, while in the region
$0.46\rho_0<\rho<0.5\rho_0$
one presumes  to have a slab--like form for the Coulomb lattice.
The parameters $R_i$ represent the finite boundary of the nucleus
and $t_i$ determines the relative surface thickness.

With these density profiles we already have a fixed proton
fraction relevant for the calculation of the pairing gap. Our calculation
of the pairing gap is a two--step process (for details, see Ref.\
\cite{elga95}).
First we solve self--consistently
the Brueckner--Hartree--Fock (BHF) equations
for the single-particle energies,
using a $G$--matrix defined through the Bethe--Brueckner--Goldstone
equation as
\begin{equation}
   G=V+V\frac{Q}{\omega -H_0}G,
\end{equation}
where $V$ is the nucleon-nucleon potential, $Q$ is the Pauli operator
which prevents scattering into intermediate
states prohibited by the Pauli
principle, $H_0$ is the unperturbed
hamiltonian acting on the intermediate
states and $\omega$ is the so--called starting energy,
the unperturbed energy
of the interacting states. Methods to solve this equation are reviewed in
Ref.\ \cite{hko95}.
The single--particle energies for state $k_i$ ($i$
encompasses all relevant
quantum numbers like momentum,
isospin projection, angular momentum, spin etc.)
in nuclear matter are assumed to have
the simple quadratic form\footnote{We set
$\hbar = c = 1$.}
\begin{equation}
   \varepsilon_{k_i}=
   {\displaystyle\frac{k_{i}^2}
   {2m^{*}}}+\delta_i ,
   \label{eq:spen}
\end{equation}
where $m^{*}$ is the effective mass.
The terms $m^{*}$ and $\delta$, the latter being
an effective single--particle
potential related to the $G$--matrix, are obtained through the
self--consistent BHF  procedure. The so--called
model--space BHF method
for the single--particle spectrum has been used, see
e.g.\ Refs.\ \cite{hko95,kuo83}, with a cutoff $k_M=3.0$ fm$^{-1}$.
This self--consistency scheme
consists in choosing adequate initial values of the
effective mass and $\delta$. The obtained $G$--matrix is in turn used to
obtain new values for $m^{*}$ and $\delta$. This procedure
continues until these parameters vary little.
The BHF equations are solved
for different proton fractions, using the formalism of Refs.\
\cite{hko95,ks93}. The nucleon--nucleon potential is defined by the
parameters of the meson--exchange potential model of the Bonn group,
version A in Table A.2 of Ref.\ \cite{mac89}.

The next step is to evaluate the gap equation following the scheme
proposed by Anderson and Morel \cite{am61} and applied
to nuclear physics
by Baldo {\em et al.\ } \cite{baldo90}. These authors introduced an
effective interaction $\tilde{V}_{k,k'}$.
This effective interaction
sums up all two--particle excitations
above the cutoff $k_M$. It is defined
according to
\begin{equation}
     \tilde{V}_{k,k'}=V_{k,k'}-\sum_{k''>k_M}V_{k,k''}\frac{1}{2E_{k''}}
                      \tilde{V}_{k'',k'},
     \label{eq:gap1}
\end{equation}
where the quasiparticle energy $E_k$ is given by
\begin{equation}
     E_k=\sqrt{\left(\varepsilon_k-\varepsilon_F\right)^2+\Delta_k^2},
     \label{eq:gap2}
\end{equation}
$\varepsilon_F$ being the single--particle energy at the Fermi surface,
$V_{k,k'}$ is the free nucleon--nucleon potential in momentum space
and $\Delta_k$ is the pairing gap
\begin{equation}
  \Delta_k=-\sum_{k'\leq k_M}\tilde{V}_{k,k'}\frac{\Delta_{k'}}{2E_{k'}}.
     \label{eq:gap3}
\end{equation}
For notational economy, we
have dropped the subscript $i$ on the single--particle energies.

In summary, first we obtain the self--consistent
BHF single--particle spectrum $\varepsilon_k$,
thereafter we solve self--consistently Eqs.\ (\ref{eq:gap1}) and
(\ref{eq:gap3}) in order to obtain the pairing gap
$\Delta$. This pairing gap is again calculated for
various proton fractions according to Eq.\ (1).
For states above $k_M$, the quasiparticle energy of
(\ref{eq:gap2}) is approximated by
$E_k=\left(\varepsilon_k-\varepsilon_F\right)$, an
approximation found to yield satisfactory results in neutron matter
\cite{elga95}.
\begin{figure}[hbtp]
     \setlength{\unitlength}{1mm}
     \begin{picture}(100,80)
     \end{picture}
     \caption{Neutron pairing energy gap $\Delta$
        as function of the position $r$
        inside the Wigner--Seitz cell for various nuclear shapes and
        densities. For the spherical phase, $\rho/\rho_0=0.058$ and
        $\rho/\rho_0=0.176$ we have used solid and dashed lines,
        respectively. The cylindrical region ($\rho/\rho_0=0.354$)
        is given by a dotted line, while the slab region
        ($\rho/\rho_0=0.48$)
        is shown with a dash-dotted line.}
     \label{fig:fig1}
\end{figure}
The results for the neutron pairing gap, effective
mass and local Fermi momentum as a function of the position in the
Wigner-Seitz cell are displayed in Figs.\ \ref{fig:fig1},
 \ref{fig:fig2} and \ref{fig:fig3}, respectively, for various density
 regions inside the crust.
\begin{figure}
     \setlength{\unitlength}{1mm}
     \begin{picture}(100,80)
     \end{picture}
      \caption{Neutron effective mass ratio $m^*/m$
         as function of the position $r$
        inside the Wigner--Seitz
         cell for the different densities and shapes
        reported in Fig.\ 1}
      \label{fig:fig2}
\end{figure}
\begin{figure}[hbtp]
     \setlength{\unitlength}{1mm}
     \begin{picture}(100,80)
     \end{picture}
     \caption{Neutron Fermi momentum as function of the position $r$
        inside the Wigner--Seitz cell for various nuclear
         shapes and densities
        as reported in Fig.\ 1}
     \label{fig:fig3}
\end{figure}
For heavy and medium heavy nuclei present in the neutron star crust, we
expect mean radii of the order $6 \sim 8$ fm.
We notice in Fig.\ \ref{fig:fig3} that the Fermi momentum for $r < 6$ fm
is close to that of the
saturation density of nuclear matter, or the central
density of $^{208}$Pb. At these densities, the gap energy in
nuclear matter is generally small \cite{elga95,baldo90}. This is also
seen in Fig.\ \ref{fig:fig1}, where the gap is less than $0.5$ MeV.
The differences between the gap energies for the various shapes of
the Coulomb lattice can be retraced to the different
effective masses in Fig.\ \ref{fig:fig2}, since the
effective masses which enter the determination of the pairing gap,
differ.
For values of $r$ between $6$ and $8$ fm, close to the Fermi surface,
where the Fermi momentum of Fig.\ \ref{fig:fig3} changes rapidly,
we see the largest variations in the effective mass
and the pairing gap, the latter reaching a peak of $2.5$  MeV for the
spherical phase. For larger values of $r$, i.e.\ outside the nucleus,
the Fermi momentum stabilizes, though there are
large differences from shape
to shape. For the spherical shape given by the lowest density
$\rho/\rho_0=0.058$ we get the lowest
value of $k_F$, whereas for the slab phase with $\rho/\rho_0=0.48$ (half
nuclear matter saturation density), we have $k_F \approx 1.3$ fm$^{-1}$.
Since the neutron $^1S_0$ pairing
gap reaches its maximum at $k_F$ between
$0.8\sim 1.0$ fm$^{-1}$ \cite{elga95,baldo90},
we see from Fig.\ \ref{fig:fig1},
that the pairing gap is at its largest for the spherical phases and for
radii larger than $8$ fm. For large values of $r$, the pairing gaps are
constant, since we have uniform neutron matter at a fixed $k_F$.

The qualitative features exhibited in
Figs.\ \ref{fig:fig1}--\ref{fig:fig3},
are similar to the results of Broglia {\em et al. } \cite{bdllp94}.
However, the pairing gaps obtained
with the effective interaction of Ref.\
\cite{bdllp94}, are larger than those obtained here.
Broglia {\em et al. }
obtain maximum pairing gaps of the order of $3.5$ MeV, whereas ours
are of the order of $2.5$ MeV. Our pairing gap for uniform matter
is close to that of Baldo {\em et al. } \cite{baldo90}, who also
employ realistic nucleon--nucleon potentials.
Moreover, in applications to finite
nuclei \cite{hko95}, the nucleon--nucleon
force used here reproduces very well e.g.\ the experimental spacing
between the $0^+$ ground state and the first excited $2^+$ state of the
tin isotopes.
The smaller energy gap in this work may in
turn have important consequences
for thermal properties of neutron stars. To see this,
we evaluate the specific heat for a system of
superfluid neutrons. Here we use the thermodynamical expression
\begin{equation}
    C_V=T\biggl({\partial S \over \partial T}\biggr)_V
\end{equation}
where $V$ is the volume of the system, $T$ is
the temperature and $S$ is the total entropy.
We obtain then the following form for the specific heat
(see Ref. \cite{bdllp94} for further details)
\begin{eqnarray}
    C_{Vn}^{sup}=&{k_B\over V_{WS}}{1\over \pi T}
    \int_{V_{sup}} drr^2\int dkk^2
    {E_{k_F}(r)\over
    cosh^2\biggl(E_{k_F}(r)/2T\biggr)} \nonumber \\
    &\times \Biggl(
      {E_{k_F}(r)\over T}-{dE_{k_F}(r)\over dT}\Biggr),
\end{eqnarray}
 where $E_{k_F}(r)$ is the local quasi--particle energy
\begin{equation}
E_{k_F}(r)=\sqrt{(\epsilon_{k}(r)-\epsilon_F)^2+\Delta^2_{k_F}(r)},
\end{equation}
$V_{WS}$ is the volume of the Wigner--Seitz
cell and $V_{sup}$ is the volume
occupied by the superfluid.
With increasing temperature the system of superfluid neutrons
exhibits a phase transition towards a
normal Fermi liquid system. The contribution
to the total neutronic specific heat inside the Wigner--Seitz cell
is written as
\begin{eqnarray}
     C_{Vn}^{norm}=&
     {k_B\over V_{WS}}{1\over \pi T^2}\int_{V_{norm}}
      drr^2\int dkk^2
     \nonumber \\ &\times{(\epsilon_{k}(r)-\epsilon_F)^2\over
     cosh^2\biggl((\epsilon_{k}(r)-\epsilon_F)/2T\biggr)},
\end{eqnarray}
where $V_{norm}$ is the volume occupied by the
normal system inside the cell.

The total neutronic specific heat can be written for all
temperatures as
\begin{equation}
     C_n=C_{Vn}^{sup}+C_{Vn}^{norm}.
\end{equation}

In order to show the relevance of our results, we compute
the total fermionic specific
heat given by
\begin{equation}
     C_T=C_n+C_e
\end{equation}
where $C_e$ is the specific heat for relativistic electrons
\begin{equation}
     C_e=\pi k_B{T\over \epsilon^e_F},
\end{equation}
and $\epsilon_F$ is the Fermi energy of the electrons \cite{bbp71}.
The large anisotropy found in the local pairing energy gap
(see Fig.\ \ref{fig:fig1}), leads
to a larger specific heat for the superfluid neutron gas
with respect to that obtained for the uniform
neutron system. This is due to the weak superfluid neutron component
inside the nucleus, see Fig.\ \ref{fig:fig1}.
This  effect yields  a larger specific heat compared
to that of the superfluid neutron matter
outside the nucleus \cite{ld95}.
\begin{figure}
     \setlength{\unitlength}{1mm}
     \begin{picture}(100,80)
     \end{picture}
       \caption{Ratio $C_{T}^{nu}/C_{T}^{u}$ as
                function of temperature $T$ for spherical
                and non--spherical phases inside the crust.
                Notations as in Fig.\ 1}
        \label{fig:fig4}
\end{figure}
In Fig.\ \ref{fig:fig4} we compare
the ratio $C_{T}(n.u.)/C_{T}(u.)$
between the total fermionic specific heat evaluated accounting
for non--uniform nuclear shapes and that obtained considering
uniform neutron matter only.
We notice that the ratio
increases dramatically moving from spherical to plate--like
nuclei, and for $T>0.008$ MeV, typical of the inner crust of a
neutron star in the first $~10^2$ yr after formation,
the non--spherical
nuclear phases give specific heats which are much larger
than that for uniform neutron matter.
This strong increase
is due to the corresponding increase in the nuclear volume
observed for non--spherical phases \cite{ldpc95}. This increase
overcomes the reduction in the anisotropy obtained in the
pairing energy gap
moving from spheres to slabs.

In summary,
our results may
have consequences  for the thermal evolution of the
star compared to models where the crust is described in terms
of uniform neutron matter only. In fact properties like
the heat diffusion
time through the crust \cite{dlpb95} may be affected.
Moreover, compared to
the results reported by Broglia {\em et al. } \cite{bdllp94}, where
the Gogny interaction was employed to obtain the pairing gap,
there is a further enhancement of the
ratio $C_{T}(n.u.)/C_{T}(u.)$.

This work has been supported by the Istituto Trentino di Cultura, Italy,
the Research Council of Norway and the NorFA (Nordic Academy for
Advanced Research).

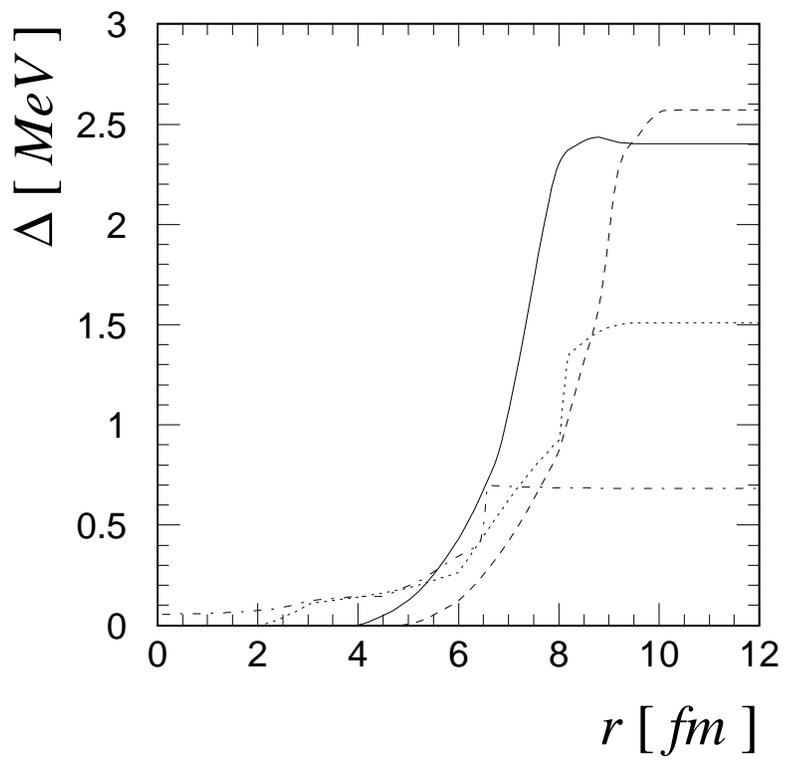
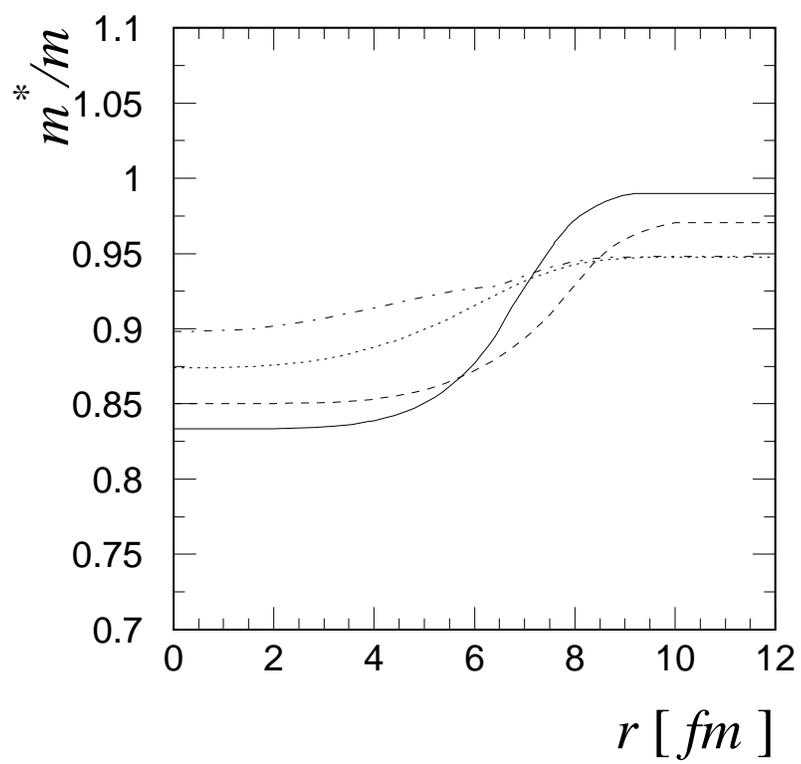
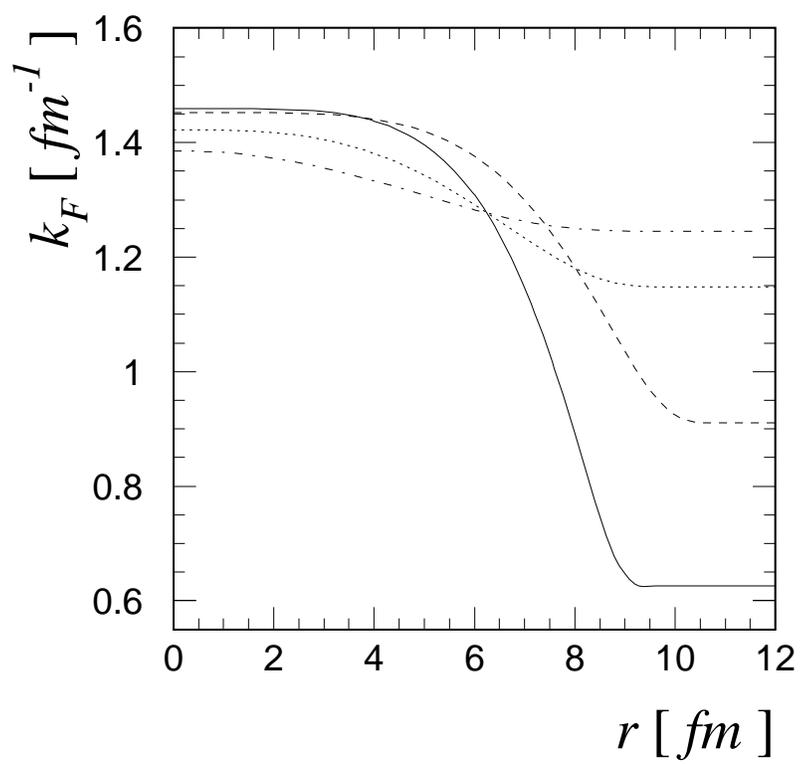
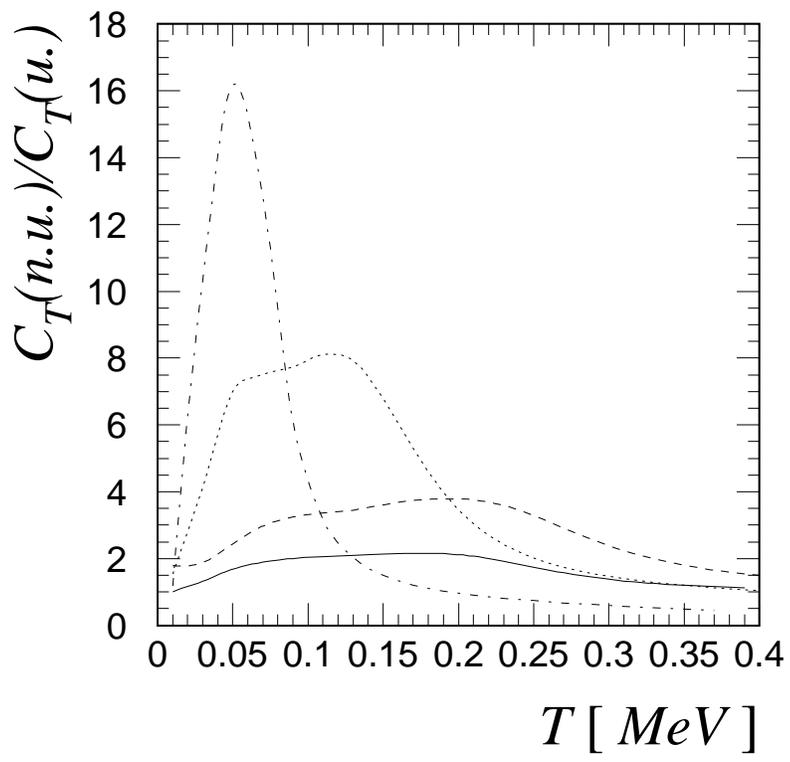
\end{document}